\pdfoutput=1
\documentclass[11pt,a4paper]{article}
\usepackage{jheppub}
\usepackage{preprintcover}  
\PreprintCoverPaperTitle{Search for a heavy top-quark partner in final states with two leptons with the ATLAS 
detector at the LHC}  
\PreprintIdNumber{CERN-PH-EP-2012-199}  
\PreprintCoverAbstract{The results of a search for direct pair production of heavy top-quark partners
in 4.7~\ifb\ of integrated luminosity from $pp$ collisions at 
$\sqrt{s}=7$~\TeV\ collected by the ATLAS detector at the LHC are reported.
Heavy top-quark partners decaying into a top quark and a neutral non-interacting 
particle  are searched for in events with two leptons in the final state.
No excess above  the Standard Model expectation is observed. 
Limits are placed on the mass of a supersymmetric \scalartop and of a spin-1/2 
top-quark partner. A spin-1/2 top-quark partner with a mass between $300 \GeV$
and  $480 \GeV$, decaying to a top quark and a neutral non-interacting particle 
lighter than $100 \GeV$, is excluded at 95\% confidence level. }  
\PreprintJournalName{JHEP}  
\usepackage{atlasphysics}
\usepackage{draftcopy}
\usepackage{lineno}
\usepackage{subfigure}
\usepackage{graphicx}
\usepackage{hyperref}
\usepackage{xspace}
\RequirePackage{lineno}

\newcommand{\sqR}   {\tilde{q}^{}_{R}}
\newcommand{\sqL}   {\tilde{q}^{}_{L}}

\newcommand{\tone}   {\tilde{t}^{}_{1}}
\newcommand{\neut}  {\tilde{\chi}^{0}_{1}}

\newcommand{\scalartop}{scalar top }     
\def\etmiss{\ensuremath{E_{\mathrm{T}}^{\mathrm{miss}}}\xspace}
\def\EtMiss{\ensuremath{E_{\mathrm{T}}^{\mathrm{miss}}}\xspace}

\renewcommand{\ttbar} {\ensuremath{t\bar{t}}\xspace}

\renewcommand{\met} {\ensuremath{E_{\mathrm{T}}^{\mathrm{miss}}}\xspace}

\def\met{\ensuremath{E_{\mathrm{T}}^{\mathrm{miss}}}}

\newcommand{\mttwo}{\ensuremath{m_{\mathrm{T2}}}\xspace}
\newcommand{\ttbarW}{\ensuremath{t\bar{t}W}\xspace}
\newcommand{\ttbarZ}{\ensuremath{t\bar{t}Z}\xspace}

\begin{document}

\title{Search for a heavy top-quark partner in final states with two leptons with the ATLAS 
detector at the LHC}

\abstract{
The results of a search for direct pair production of heavy top-quark partners
in 4.7~\ifb\ of integrated luminosity from $pp$ collisions at 
$\sqrt{s}=7$~\TeV\ collected by the ATLAS detector at the LHC are reported.
Heavy top-quark partners decaying into a top quark and a neutral non-interacting 
particle  are searched for in events with two leptons in the final state.
No excess above  the Standard Model expectation is observed. 
Limits are placed on the mass of a supersymmetric \scalartop and of a spin-1/2 
top-quark partner. A spin-1/2 top-quark partner with a mass between $300 \GeV$
and  $480 \GeV$, decaying to a top quark and a neutral non-interacting particle 
lighter than $100 \GeV$, is excluded at 95\% confidence level. 
}

\maketitle

\section{Introduction}

Partners of the top quark are an ingredient of several
models addressing the hierarchy problem of the Standard Model (SM). 
In order to stabilize the Higgs boson mass against divergent quantum corrections,
these new particles should have masses close to the electroweak symmetry 
breaking energy scale, and thus be accessible at the LHC.
One of these models is 
Supersymmetry (SUSY) ~\cite{Miyazawa:1966,Ramond:1971gb,Golfand:1971iw,Neveu:1971rx,Neveu:1971iv,Gervais:1971ji,Volkov:1973ix,Wess:1973kz,Wess:1974tw} which naturally
resolves the hierarchy problem \cite{Weinberg:1975gm,Gildener:1976ai,Weinberg:1979bn,Susskind:1978ms} by introducing supersymmetric partners of the known
bosons and fermions. In the MSSM~\cite{Fayet:1976et,Fayet:1977yc,Farrar:1978xj,Fayet:1979sa,Dimopoulos:1981zb}, an R-parity conserving minimal supersymmetric extension of the SM, 
the scalar partners of right-handed and left-handed quarks,
$\sqR$ and $\sqL$, can mix to form two mass eigenstates.
In this paper a search for a \scalartop $\tone$
which decays into a top quark and the lightest neutralino $\neut$ 
is performed. In this model, the $\neut$ is a stable particle which 
would escape detection. \par

A top-quark fermionic partner $T$ which decays into a stable, neutral, weakly interacting 
particle $A_0$ also appears in other SM extensions, such as  
little Higgs models
with $T$-parity conservation \cite{Cheng:2003ju,Cheng:2004yc,Cheng:2005as} 
or models of Universal Extra Dimensions (UED) 
with Kaluza-Klein parity \cite{Appelquist:2000nn}. 
The production cross section at the LHC 
is predicted to be approximately six times higher for 
fermionic $T$~\cite{Alwall:2010jc}  than for the $\tone$.  
Furthermore, \scalartop and $T$ decay kinematic distributions differ due to 
helicity effects in the decay, yielding different experimental acceptances.\par

Searches for these spin-1/2 heavy top-quark partners were performed by the CDF
Collaboration in proton-antiproton collisions at $\sqrt{s}=1.96 \TeV$~\cite{Aaltonen:2011na}, 
excluding at 95\% confidence level (CL) top-quark partners with masses up to $400~\GeV$
for an $A_0$ mass lower than $70 \GeV$. A previous ATLAS analysis 
with 1.04~\ifb $\;$ of proton-proton collisions at $\sqrt{s}=7 \TeV$~\cite{Aad:2011wc} 
excludes a $T$ with masses up to $420~\GeV$.

In this paper a search for the direct pair production of  heavy top-quark partners 
is presented, where $\tone \rightarrow t \ninoone$ or $T \rightarrow t A_0$. 
The  final state targeted by the analysis includes two top quarks and additional missing transverse 
momentum $\mathbf p_{\mathrm{T}}^{\mbox{\footnotesize miss}}$,  with magnitude
$\etmiss$, resulting mainly from the undetected $\neut$ or $A_0$.
The present study addresses the two-lepton signature resulting from the leptonic
decay of both the $W$ bosons from the top-quark decay. The neutrinos from the 
$W$ decays also contribute to the missing transverse momentum. 
Events with two electrons, two muons, or an electron-muon pair in the 
final state are selected by the analysis.
To separate the signal from the large irreducible background from  top-quark 
pair production, the $\mttwo$ variable \cite{Lester:1999tx,Barr:2003rg} is used.
It is defined as:

\begin{equation*} 
\mttwo
( \mathbf p^{\ell_1}_{\mathrm{T}}, \mathbf p^{\ell_2}_{\mathrm{T}}, \mathbf
p^{\mbox{{\footnotesize miss}}}_{\mathrm{T}}) = \min_{\mathbf q_{\mathrm{T}} + \mathbf r_{\mathrm{T}} = 
\mathbf p^{\mathrm{miss}}_{\mathrm{T}} } 
\left\{ \max [\; 
    m_{\mathrm{T}}( \mathbf p^{\ell_1}_{\mathrm{T}}, \mathbf q_{\mathrm{T}} ),  
    m_{\mathrm{T}}( \mathbf p^{\ell_2}_{\mathrm{T}}, \mathbf r_{\mathrm{T}} 
    ) 
\;] 
\right\} ,
\end{equation*}

where $m_{\mathrm T}$ indicates the transverse mass, $\mathbf p^{\ell_1}_{\mathrm{T}}$ and
$\mathbf p^{\ell_2}_{\mathrm{T}}$ are the transverse momenta of the two leptons, and $\mathbf
q_{\mathrm{T}}$ and $\mathbf r_{\mathrm{T}}$ are
vectors which satisfy
$\mathbf q_{\mathrm{T}} + \mathbf r_{\mathrm{T}} = \mathbf p^{\mbox{\footnotesize miss}}_{\mathrm{T}}$. 
The minimization is performed over all the possible decompositions 
of  $\mathbf p^{\mbox{\footnotesize miss}}_{\mathrm{T}}$.
The distribution of this variable presents a sharp kinematic limit at the
$W$ boson mass for \ttbar production \cite{Cho:2007dh,Burns:2008va}, whereas
for the signal topology it decreases slowly towards a 
higher mass value, due to the presence of the two additional
invisible particles produced in association with the top-quark pair.
The results are interpreted in the scalar top$-$neutralino mass plane
as well as in a generic model producing
a heavy spin-1/2 top-quark partner $T$ decaying into an
invisible particle $A_0$ and a top quark.\par
This analysis is sensitive to masses of the
top-quark partner in excess of about $200 \GeV$ and 
is thus complementary to a parallel ATLAS study reported in refs.~\cite{Aad:2011xx,Aad:2011yy} optimized 
for  \scalartop masses near or below the top mass.

\section{The ATLAS detector}

The ATLAS detector~\cite{Aad:2008zzm} consists of inner tracking devices
surrounded by a superconducting solenoid, electromagnetic and 
hadronic calorimeters and a muon spectrometer with a toroidal magnetic
field. The inner detector, in combination with the axial 2~T field from the 
solenoid, provides precision tracking of charged particles for 
$| \eta | < 2.5$, where the pseudorapidity $\eta$ is defined in terms of the 
angle $\theta$ with the beam pipe axis as  $\eta = -\ln \tan(\theta/2)$.
It consists of a silicon pixel detector, a 
silicon strip detector and a straw tube tracker that also provides
transition radiation measurements for electron identification.
The calorimeter system covers the pseudorapidity range $| \eta | < 4.9$.
It is composed of sampling calorimeters with either liquid argon 
or scintillating tiles as the active media. The muon spectrometer has 
separate trigger and high-precision tracking chambers which provide 
muon trigger and measurement capabilities for $| \eta | < 2.4$ and 
$|\eta| < 2.7$ respectively.

\section{Monte Carlo samples}
\label{sec:MC}

Monte Carlo (MC) simulated event samples are used to aid in the description of the background 
and to model the SUSY and spin-1/2 heavy top-quark partner signals. 

Top-quark pair and $Wt$ production are simulated with 
{\sc mc@nlo}~\cite{Frixione:2002ik,Frixione:2005vw}, interfaced with
{\sc herwig}~\cite{Corcella:2000bw} for the fragmentation and 
the hadronization processes, including {\sc jimmy}~\cite{Butterworth:1996zw} 
for the underlying event. The top-quark mass is fixed at $172.5 \GeV$,   
and the next-to-leading-order (NLO) parton distribution function (PDF) set
CTEQ10~\cite{Lai:2010vv} is used. Additional MC samples are used to
estimate the event generator systematic uncertainties: 
two {\sc powheg}~\cite{Frixione:2007vw} samples,
one interfaced with {\sc herwig} and the other
with {\sc pythia}~\cite{Sjostrand:2006za}; an {\sc alpgen}~\cite{Mangano:2002ea}
sample interfaced with {\sc herwig} and {\sc jimmy}; two 
{\sc acermc}~\cite{Kersevan:2004yg} samples produced with
variations to the {\sc pythia} parton shower parameters chosen  
such that the two samples produce additional radiation consistent with the 
experimental uncertainty in the data~\cite{ATLAS:2012al,ATLAS-CONF-2011-009}.

Samples of $Z/\gamma^\star$ produced in association with light- and 
heavy-flavour jets
are generated with {\sc alpgen} using the PDF set CTEQ6.1~\cite{Pumplin:2002vw}. 
Samples of \ttbarZ and \ttbarW production are generated with {\sc madgraph}~\cite{Alwall:2011uj} 
interfaced to {\sc pythia}. Diboson samples ($WW$, $WZ$, $ZZ$) 
are generated with {\sc sherpa}~\cite{Gleisberg:2008ta}. Additional samples generated
with {\sc alpgen} and 
{\sc herwig} are used for the evaluation of the event generator systematic uncertainties. 

The background predictions are normalized to theoretical cross sections, including 
higher-order QCD corrections when available, and are compared to data in control 
regions populated by events produced by SM processes.
Next-to-next-to-leading-order (NNLO) cross sections are used for 
inclusive $Z$ boson production~\cite{Hamberg:1990np,Gavin:2012kw}.
Approximate NLO+NNLL (next-to-next-to-leading-logarithms) cross 
sections are used in the normalization of the \ttbar~\cite{Aliev:2010zk} and 
$Wt$~\cite{Kidonakis:2010ux} samples. NLO cross sections are used for the 
diboson samples~\cite{Frixione:2002ik,Binoth:2006mf} and for the \ttbarW
and \ttbarZ~\cite{Lazopoulos:2008de} samples.  Production of  \ttbar  in 
association with \bbbar$\;$ is normalized to leading 
order (LO) cross section~\cite{Mangano:2002ea}. 
Table~\ref{tab:xsec} summarizes the production cross sections used in this
analysis and their uncertainties. 

\begin{table}
\begin{center}
\caption{The most important SM background processes and their production
cross sections, multiplied by the relevant branching ratios. The $\ell$ indicates
all three types of leptons ($e, \mu, \tau$) summed together. The $Z/\gamma^\star$ production
cross section is given for events with a di-lepton invariant mass of at least $12 \GeV$. }
\label{tab:xsec}
\vspace{0.5cm}
\begin{tabular}{|l|c|c|}
\hline
Physics process & $\sigma \cdot$BR [pb] & Perturbative order \\
\hline
$Z/\gamma^\star \rightarrow \ell^+ \ell^-$ &  $1069  \pm 53$             & NNLO \\
\ttbar                                                                     & $167^{+17}_{-18}$          & NLO+NNLL \\
$Wt$                                                           & $15.7 \pm 1.2$& NLO+NNLL \\                                 
\ttbarW                                                        & $0.168^{+0.023}_{-0.037}$    & NLO \\
\ttbarZ                                                         & $0.130 \pm 0.019$    & NLO \\
$WW$                                            & $44.4 \pm 2.8$    & NLO \\
$WZ$                                             & $19.1 \pm 1.3$    & NLO \\
$ZZ$                                              & $ 6.2 \pm 0.3$    & NLO \\
\hline
\end{tabular}
\end{center}
\end{table}

SM processes that generate jets which are misidentified as leptons, 
or where a lepton from a $b$-hadron or $c$-hadron decay is selected, collectively
referred to as ``fake'' leptons in the following, are estimated from data
as described in section~\ref{sec:backgrounds}. 

Scalar top signal samples are generated with  {\sc Herwig++}\cite{Bahr:2008pv}. 
The mixings in the scalar top and gaugino sector are chosen to 
be such that the lightest scalar top is mostly the partner $\tilde t_{\mathrm R}$
of the right-handed top quark, and the lightest neutralino
is almost a pure bino. Under such conditions, the scalar top is 
expected to decay to the lightest neutralino and a top quark with a branching 
ratio close to 100\%, even if the decay mode to a chargino and a $b$ quark
is kinematically allowed. The effects of helicity in the decay are correctly treated 
by {\sc Herwig++}. Spin-1/2 heavy top-quark partner signal samples are 
generated with {\sc madgraph}~\cite{Alwall:2011uj}.
Signal cross sections are calculated to NLO 
in the strong coupling constant, including the resummation of soft gluon
emission at next-to-leading-logarithmic accuracy
(NLO+NLL) \cite{Beenakker:1997ut,Beenakker:2010nq,Beenakker:2011fu},
as described in ref.~\cite{Kramer:2012kl}.

The MC generator parameters have been tuned to ATLAS data ~\cite{tune1,tune2}  and generated events
have been processed through a detector simulation~\cite{Aad:2010ah} based 
on {\sc geant4}~\cite{Agostinelli:2002hh}. Effects of multiple proton-proton interactions 
in the same bunch crossing (pile-up) are included, with the MC samples re-weighted so that the 
distribution of the average number of interactions per bunch crossing agrees with that in the data.

\section{Physics object reconstruction}
\label{sec:objdef}

Proton-proton interaction vertex candidates are reconstructed using
the Inner Detector tracks. The vertex with the highest scalar sum of the $\pT$
of the associated tracks is defined as the primary vertex. 

Jets are reconstructed from three-dimensional calorimeter energy clusters 
using the anti-$k_t$ jet algorithm~\cite{Cacciari:2005hq,Cacciari:2008gp} 
with a radius parameter of 0.4. The measured jet energy is corrected for 
inhomogeneities, and the non-compensating nature of the calorimeter 
with $\pT$- and $\eta$-dependent correction factors~\cite{Aad:2011he}.
Only jet candidates with $\pT > 20 \GeV$, $|\eta| < 2.5$ and a ``jet vertex
fraction" larger than 0.75 are retained. Based on tracking information, the 
jet vertex fraction quantifies the fraction of a jet's momentum that originates
from the reconstructed primary vertex. The requirement on the jet vertex fraction
 rejects jets originating from additional proton-proton interactions occurring 
 in the same bunch crossing. Events with 
any jet that fails the jet quality criteria designed to reject noise and non-collision 
backgrounds~\cite{Aad:2011he} are rejected. 

Electron candidates are required to have $\pT > 20 \GeV$, $|\eta| < 2.47$ and 
to satisfy ``medium" electromagnetic shower shape and track selection  
quality criteria~\cite{Aad:2011mk}. These preselected electrons
are then required to pass ``tight'' quality
criteria~\cite{Aad:2011mk} which places additional requirements on the ratio of 
calorimetric energy to track momentum, and on the fraction of hits in the straw
tube tracker that pass a higher threshold for  transition radiation.
The electron candidates are then required to be isolated: the scalar sum of the $\pT$,
$\Sigma \pT$, 
of inner detector tracks, not including the electron track, with $\pT > 1 \GeV$ within
a cone in the $\eta - \phi$ plane of radius 
$\Delta R = \sqrt{\Delta\eta^2+\Delta\phi^2} = 0.2$ around the electron candidate must be less 
than 10\% of the electron \pT.

Muon candidates are reconstructed using either a full muon spectrometer track
matched to an inner detector track, or a muon spectrometer segment matched
to an extrapolated inner detector track~\cite{ATLAS-CONF-2011-063}. 
They must be reconstructed with sufficient hits in the pixel, strip and straw tube 
detectors. They are required to have 
$\pT > 10 \GeV$, $|\eta| < 2.4$ and must have longitudinal and transverse 
impact parameters within 1~mm and 0.2~mm of the primary vertex, respectively. 
Such preselected candidates are then 
required to have $\Sigma \pT < 1.8 \GeV$, defined in analogy to the electron
case.

Following the object reconstruction described above, overlaps between jet, 
electron and muon candidates are resolved. Any jet within $\Delta R = 0.2$ 
of preselected electrons is discarded. Electrons or muons within $\Delta R = 0.4$ 
of any remaining jet are then discarded to reject leptons from the decay 
of a $b$- or $c$-hadron.

The  \EtMiss is  the magnitude of the vectorial sum 
of the $\pT$ of the reconstructed jets (with $\pT > 20 \GeV$ and $|\eta| < 4.5$)
after overlap removal, preselected leptons  and clusters of calorimeter cells not
belonging to reconstructed physics objects~\cite{Aad:2012re}. 

A $b$-tagging algorithm exploiting both impact parameter and secondary 
vertex information~\cite{btag} is used to identify jets containing a $b$-hadron 
decay. The chosen operating point has a 60\% efficiency for tagging $b$-jets in a MC sample
of \ttbar events, with a mis-tag probability of less than 1\% for jets from 
light quarks and gluons.

\section{Event selection}

This search uses proton-proton collisions recorded in 2011 at a centre-of-mass 
energy of 7~\TeV. The data are selected with a three-level trigger system. 
Events are accepted if they pass either
a single-electron trigger reaching a plateau efficiency of about 97\%
for electrons with $\pT > 25 \GeV$, or a single-muon or combined 
muon+jet trigger which reaches a plateau efficiency of about 75\% (90\%) 
in the barrel (end-caps) for events including muons with $\pT > 20 \GeV$
and jets with $\pT > 50 \GeV$. The combined muon+jet trigger is 
used for the data-taking periods with high instantaneous luminosity,
because it is based on looser muon identification requirements than the 
single-muon trigger available for those periods, resulting in a higher plateau efficiency.
Events are required to have a reconstructed primary vertex with five or more
tracks consistent with the transverse beam spot position.
Following beam, detector and data quality requirements, a 
total integrated luminosity of $(4.7 \pm 0.2)$~\ifb is used, measured as described 
in refs.~\cite{lumi1,Aad:2011dr}.

Two signal regions (SRs) are defined, one for different-flavour, and one 
for same-flavour leptons. For both SRs 
events are required to have exactly two opposite-sign (OS) leptons (electrons or 
muons) with an invariant mass larger than $20 \GeV$. 
At least one electron or muon must have a momentum in the trigger efficiency
plateau regions described above.
If the event contains a third 
preselected electron or muon, the event is rejected. At least 
two jets with $\pT > 25 \GeV$, and at least one of them 
with $\pT > 50 \GeV$, are required. This requirement suppresses $WW$
and $Z/\gamma^\star$+jets$\;$ backgrounds.\par 
For the same-flavour SR, additional selections are imposed to suppress the 
$Z/\gamma^\star$+jets, $WZ$ and $ZZ$ backgrounds, which represent
a significant fraction of events with large \mttwo.  These events 
have large $\met$, which for the $WZ$ process is generated  
by the leptonic decay of the $W$ boson, for the $ZZ$ process by the 
decay of one of the $Z$ bosons to neutrinos, and for $Z/\gamma^\star$+jets 
by the tails in the jet energy resolution. The additional selections required 
in the same-flavour channel to suppress these backgrounds are that 
the invariant mass of the leptons must 
be outside the $71-111 \GeV$  range, and at least one of the jets must
be tagged as a $b$-jet. After these selections the background is dominated by \ttbar.

Finally, for both SRs, signal candidate events are required to have a value of \mttwo
larger than $120 \GeV$. This requirement suppresses the remaining \ttbar 
and $WW$ backgrounds by several orders
of magnitude and was chosen to  optimize the coverage of the 
analysis in the $\tone-\neut$ and $T-A_0$ planes.

\begin{table}
\begin{center}
\caption{Efficiency of the \mttwo selection, calculated after all other selection requirements
applied in the SR, for signal samples with different values of the mass of the 
\scalartop or of the spin-1/2 heavy top-quark partner. The mass of the $\ninoone$ or $A_0$ is zero in each
case.  No signal sample with mass $m(T) = 200 \GeV$ has been simulated.}
\vspace{0.5cm}
\label{tab:mt2eff}
\begin{tabular}{|l|c|c|c|c|c|}
\hline
Top quark partner mass [\GeV ] & 200 & 300 &  400 & 500 & 600 \\
\hline
$\tone \tone$ production  & 0.02\% & 7.7\% & 22.0\% & 35.6\% & 43.0\% \\
$TT$ production                &     -         & 5.3\% & 15.8\% & 27.3\% & 34.3\% \\
\hline
\end{tabular}
\end{center}
\end{table}

Before the $\mttwo$ selection, $\ttbar$ production is by far the largest background. 
The efficiency of the $\mttwo$ selection for \ttbar events, calculated after all the other 
SR cuts, is 0.007\%.
The efficiency of the \mttwo selection for \scalartop and spin-1/2 heavy top-quark partner 
signal samples is given in table~\ref{tab:mt2eff} for several values of the top-quark 
partner mass and for a massless $\ninoone$ or $A_0$. The efficiency is smallest when 
$\Delta m = m(\tone)-m(\ninoone)$ or $m(T)-m(A_0)$ is close to the top quark mass, 
because the kinematics of the signal are then similar to those of \ttbar background, and 
it increases with increasing $\Delta m$. For equal masses, the spin-1/2 top-quark partner signals
have a slightly lower efficiency than the \scalartop signals, due to helicity 
effects in the decay.

\section{Background estimation}
\label{sec:backgrounds}

The dominant SM background contributions to the 
SRs are top-quark pair and $\Zboson/\gamma^\star$+jets
production.
They are evaluated by defining a control region (CR)  populated 
mostly by the targeted background, and using MC simulation to extrapolate
from the rate measured in the CR to the expected background 
yield in the SR:

\begin{displaymath}
N(\mathrm{SR}) = \left( N^{\mbox{\tiny Data}}(\mathrm{CR}) -  N_{\mbox{\tiny others}} (\mathrm{CR}) \right)
                \frac{N^{\mbox{\tiny MC}}(\mathrm{SR})}{N^{\mbox{\tiny MC}}(\mathrm{CR})} 
\end{displaymath}

\noindent where $N^{\mbox{\tiny Data}}(\mathrm{CR})$ is the number of 
data events observed in the CR, $N^{\mbox{\tiny MC}}(\mathrm{CR})$ and 
$N^{\mbox{\tiny MC}}(\mathrm{SR})$ are the number of events of the targeted 
background expected 
from MC simulation in the CR and SR respectively, and 
the term $N_{\mbox{\tiny others}}(\mathrm{CR})$ is the contribution
from the other background sources in the CR which is estimated from MC 
simulation or additional control samples in data. 
The ratio of the number of MC events in the SR to the number of MC 
events in the CR 
for a given background source is referred to as the 
``transfer factor'' in the following.

\begin{table}
\begin{center}
\caption{Expected background composition and comparison of the 
predicted total SM  event yield to the observed number 
of events in the top-quark control regions described in the text for the 
same-flavour (SF) and different-flavour (DF) selections.  The expected
\Zboson$/\gamma^\star$+jets rate in the DF channel is negligible. 
The quoted uncertainties include the systematic uncertainties 
described in section~\ref{systematics}.}
\vspace{0.5cm}
\label{tab:ttbarCR}
\begin{tabular}{|l|c|c|}
\hline
                & $\ttbar$ CR & $\ttbar$ CR\\
Process & DF & SF \\
\hline
\ttbar                          &    $68 \pm 11$     & $39 \pm 11$ \\
\ttbarW + \ttbarZ       & $0.37 \pm 0.07$ & $0.20 \pm 0.05$ \\
$Wt$                          & $2.7  \pm 1.0$     & $1.8 \pm 0.6$ \\    
$Z/\gamma^\star$+jets &  -   & $3.5 \pm 1.4$ \\
Fake leptons           & $0.4 \pm 0.3$       & $0.5 \pm 1.6$ \\
Diboson                   & $0.49 \pm 0.14$  & $0.10 \pm 0.05$ \\
\hline
Total non-$\ttbar$      & $4.0 \pm 1.5$       & $6.1 \pm 3.7$ \\
Total expected          & $72 \pm 11$         & $45 \pm 12$ \\ 
\hline
Data                    &       79            &        53            \\
\hline
\end{tabular}
\end{center}
\end{table}

The \ttbar CR is defined akin to the SR, except for 
\mttwo, which is required to be between 85~GeV and 100~GeV.
The expected background composition of the $\ttbar$ CR is given
in table~\ref{tab:ttbarCR}. The contamination due to fake leptons is 
evaluated from data with 
the technique described below, while all the other processes 
are obtained from the MC prediction. The \ttbar background is expected to
account for 86\% and 94\% of the SM rate in the same-flavour and 
different-flavour CRs, respectively.
The number of observed events is in good agreement with 
the expected event yields. 

The systematic uncertainties on the modelling of the \ttbar background 
transfer factor due to the choice of the MC generator are assessed
by comparing the baseline sample simulated with 
{\sc mc@nlo} with the alternative samples described in 
section~\ref{sec:MC}.

The background from $Z/\gamma^\star$+jets is relevant for the 
same-flavour selection in the case of the decay channels $Z\rightarrow ee$ or $\mu\mu$.
For $Z\rightarrow \tau\tau$ decays, which would contribute both to 
the same-flavour and the different-flavour samples, the \mttwo
distribution  falls very steeply, and the number of expected events for
\mttwo in excess of $80 \GeV$ is negligible.

The CR for $Z/\gamma^\star$+jets
is defined with the same selections as for the SR, except for the $Z$ boson veto selection 
which is reversed. The observed number of events in this
CR is 11, compared to 7.6$\pm$1.1 expected, of which 7.0$\pm$1.1
are from $Z$ boson production, where the quoted uncertainties include the 
systematics discussed in the next section.
The transfer factor between CR and SR is evaluated with $Z/\gamma^\star$+jet  MC
samples to which all the selections of the same-flavour analysis except the 
$b$-tagging requirement are applied. Detailed checks have been performed
in order to verify that this transfer factor, which relates the 
number of events inside the $Z$ boson peak to the number of events
outside, is stable with respect to the \mttwo and $b$-tag requirements. 
The method is validated using an auxiliary CR dominated by 
$Z/\gamma^\star$+jets events, defined in the same way as the SR except  
the $b$-jet requirement is removed. The number of predicted 
background events is 7.5$\pm$1.3 (of which 7.2$\pm$1.3  
from $Z/\gamma^\star$+jets) while the observed number is 10. 
The quoted uncertainty on the prediction is  only  statistical. 

Additional SM processes yielding  two isolated 
leptons and \EtMiss ($W t$, $WW$, $WZ$,
$ZZ$, $\ttbarW$, $\ttbarZ$) are estimated from 
the MC simulation.
The contribution from diboson processes, particularly $WW$ production, is about
a quarter of the total background in the different-flavour signal region. 
The high-\mttwo population in $WW$ production is 
dominated by events in which a strongly off-shell $W$ is produced. 
The 18\% relative difference in the number of events with $\mttwo>120$~GeV
between {\sc sherpa} and {\sc herwig} before jet selection 
is taken as a systematic uncertainty on the simulation of the \mttwo distribution. 
The 45\% relative
difference between {\sc sherpa} and  {\sc alpgen} in the efficiency of the jet 
selections integrated over the whole \mttwo range is taken as a  
systematic uncertainty on the $WWjj$ cross section. 

The fake lepton background consists of semi-leptonic  $\ttbar$,
$s$-channel and $t$-channel single top, 
$W$+jets and light- and heavy-flavour jet production.
The contribution from this  background is small (less than 10\% of the total 
background). It is estimated from data 
with a method similar to that described in refs.~\cite{top1,top2}.
Two types of lepton identification criteria are defined for 
this evaluation: ``tight'', corresponding to the full set of identification
criteria described above, and ``loose''  corresponding to preselected
electrons and muons. 
The method counts the number of observed events containing loose-loose, 
loose-tight, tight-loose and tight-tight lepton pairs in the SR.
The probability for real leptons passing the loose selection criteria to also pass the tight selection 
is measured using a $Z\to\ell \ell$ sample. The equivalent probability for fake leptons is measured
from multijet-enriched  control samples.
From these  probabilities the number of 
events containing a contribution from one or two fake leptons is calculated. 

The procedure described above is used to estimate the fake lepton background 
with looser selections and extrapolated to the signal region. A systematics 
uncertainty is assigned to the extrapolation procedure by comparing the 
direct and extrapolated background estimate in various control regions.

\par

\section{Systematic Uncertainties}
\label{systematics}

Various systematic uncertainties affecting the predicted background rates
in the signal regions are considered. Such uncertainties are either used directly
in the evaluation of the predicted background in the SR (for diboson, $Wt$, 
\ttbarW and \ttbarZ production), or to compute the uncertainty on the transfer 
factor and propagate it to the predicted event yields in the SR (for \ttbar, 
$Z/\gamma^{\star}$+jets).

The following experimental systematic uncertainties were found to be 
non-negligible:

{\bf Jet energy scale and resolution.} The uncertainty on the jet energy 
scale (JES), derived using single particle response and test beam data, 
varies as a function of the jet $\pT$ and pseudorapidity~\cite{Aad:2011he}. 
Additional systematic 
uncertainties arise from the dependence of the jet response on the number
of interactions per bunch crossing and on the jet flavour. The total 
jet energy scale uncertainty at $\pT = 50 \GeV$ in the central detector region
is about 5\%~\cite{Aad:2011he}. The components of the jet energy scale 
uncertainty are varied by $\pm 1 \sigma$ in the MC simulation in order to obtain 
the resulting uncertainty in the event yield. Uncertainties related to the jet 
energy resolution (JER) are obtained with an in situ measurement of 
the jet response asymmetry in dijet events~\cite{jer}. Their impact on the
event yield is estimated by applying an additional smearing to the jet 
transverse momenta. The JES and JER variations applied to the jet momenta are 
propagated to the \EtMiss. The JES and JER relative uncertainties on the 
same-flavour and different-flavour signal region event yield amount to 
16\% and 22\%, respectively.

{\bf Calorimeter cluster energy scale and pile-up modelling.} The 
uncertainties related to the contribution to \EtMiss from the energy in the 
calorimeter cells not associated to electrons, muons or jets, and also from low 
momentum ($7 \GeV < \pT < 25 \GeV$) jets, as well as the uncertainty due to the 
modelling of pile-up have been evaluated to amount to 6\% (25\%) of the 
same-flavor (different-flavor) event yield. The fractional uncertainty is 
smaller in the same-flavour channel
because it has a very small impact (2\%) on the estimation of the $Z/\gamma^\star$+jets 
background, which is by far the largest contribution to the same-flavour channel. 

\boldmath$b${\bf-tagging efficiency and mis-tagging uncertainties.} \unboldmath
This uncertainty is evaluated
by varying the $b$-tagging efficiency and mis-tagging rates within the 
uncertainties measured in situ~\cite{btag}. 
Since the different-flavour selection does not make use of $b$-tagging, this uncertainty
only affects the same-flavour channel and is relatively small (about 1\% of the total 
event yield). 

{\bf Fake-lepton background uncertainties :}  An uncertainty of 
 33\% (25\%) is assigned to the fake background in the same-flavour 
 (different-flavour)  channel
 from the comparison of results from different CRs, with an additional 30\% 
 is taken as the systematic uncertainty due to the projection of events into the SR. 

 \vskip\baselineskip 

Other significant sources of uncertainty are the normalization 
uncertainties for processes estimated from MC simulation only, the theoretical uncertainties
discussed in section~\ref{sec:backgrounds}, the limited number of data events 
 in the CRs, the limited number of MC events, and the integrated luminosity.

 A summary of the uncertainties on 
 the total expected background in the two channels is given in 
 table~\ref{tab:uncertainties}. The row labelled ``statistics'' includes
 the effects of the limited number of data events in the CRs and the limited number of
 MC events.
 The theoretical uncertainties include the cross section, MC 
 generator, and initial- and final-state radiation uncertainties. They are 
 smaller for the same-flavor channel because the theoretical uncertainty 
 on the $Z/\gamma^\star$+jets is relatively small (about 10\%). In the 
 opposite-flavour channel the dominant backgrounds are the 
 top pair production, which is affected by  an uncertainty of about 100\%
 (due to the description of the high \mttwo tail in simulation), and the diboson
 process whose cross section has an uncertainty of about 50\%, due mostly
 to the poor prediction of the production in association with two (or more)
 jets.

\begin{table} [htb!]
\caption{Total expected background yield and uncertainties 
in the same-flavour (SF) and different-flavour (DF) signal regions.  
Where the uncertainty is not symmetric, the upwards and downwards values
are given.}
\label{tab:uncertainties}
\begin{center}
\begin{tabular}{|l|c|c|c|c|c|c|c|}
\hline
\bf{Channel}       &     \bf{SF}        &      \bf{DF}     \\
\hline
Total event yield    &    1.58       &          0.94    \\
\hline
JES + JER    &      16\%    &         22\%  \\
$b$-tagging &       1\%      &          --      \\
\EtMiss and pile-up modeling      &      6\%       &          25\%  \\
Luminosity     &     1\%       &             2\%   \\
Theory            &    14\%     &           48\%  \\
Statistics        &    $\pm^{29}_{26}$\%     &           20\%  \\
Fake-lepton uncertainties  &  $\pm^8_0$\%        &     $\pm^9_0$\%    \\ 
\hline
Total uncertainty      &    $\pm^{40}_{37}$\%     &            64\%  \\
\hline
\end{tabular}
\end{center}
\end{table}

For the limit calculation, the uncertainty on the expected signal yield is also needed.
The JES, JER, calorimeter energy scale and event pileup, and $b$-tagging uncertainties
discussed above have been taken into account. The typical total uncertainty from these sources 
varies between 4\% and 12\% for the DF channel and between 7\% and 22\%
for the SF channel,
with comparable contributions from the JES, the calorimeter energy scale and pileup, and 
(for SF) the $b$-tagging.uncertainties.

The uncertainty on the signal cross sections 
is calculated with an envelope of cross section predictions which is defined 
using the 68\% confidence level (CL) ranges of the CTEQ~\cite{Nadolsky:2008zw}
(including the $\alpha_S$ uncertainty) and MSTW~\cite{Martin:2009iq}
PDF sets, together with  variations of the factorization and
renormalization scales by factors of two or one half. The nominal
cross section value is taken to be the midpoint of the envelope and
the uncertainty assigned is half the full width of the envelope,
following the PDF4LHC recommendations~\cite{Botje:2011sn}
and using the procedure described in ref.~\cite{Kramer:2012kl}. The typical cross 
section uncertainty is 12\% for the spin-1/2 top-quark partner signal and 15\%
for the \scalartop signal.

\section{Results}
\label{sec:results}

Figure~\ref{fig:mt2} shows the distributions of the \mttwo variable 
for same-flavour and different-flavour events after all
selection criteria are applied except the selection on \mttwo itself. 
For illustration, the distributions for two 
signal hypotheses are also shown. The data agree with the
SM background expectation within uncertainties. 

\begin{figure}[htb!]
\begin{center}
\includegraphics[width=0.7\textwidth]{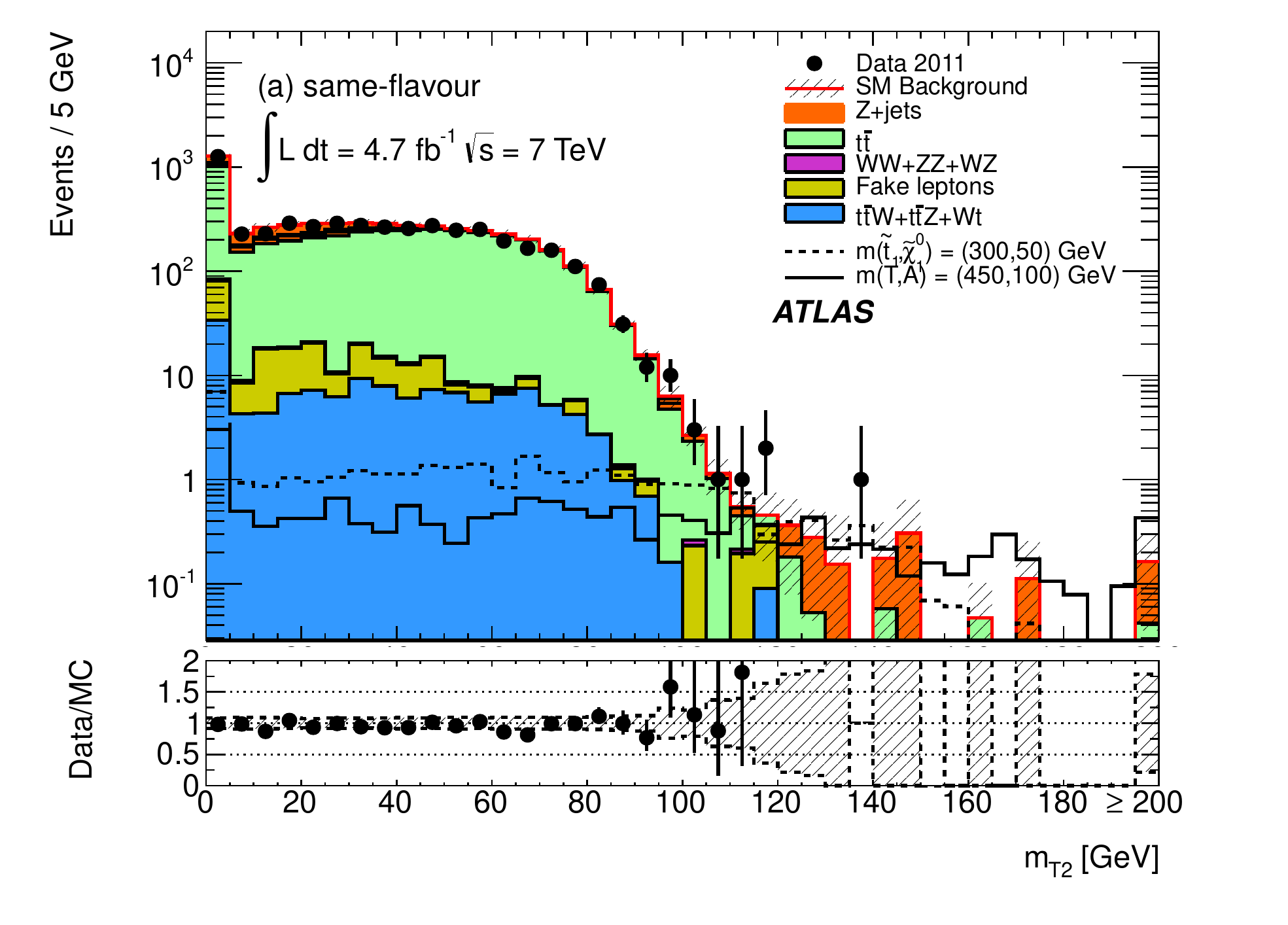}
\includegraphics[width=0.7\textwidth]{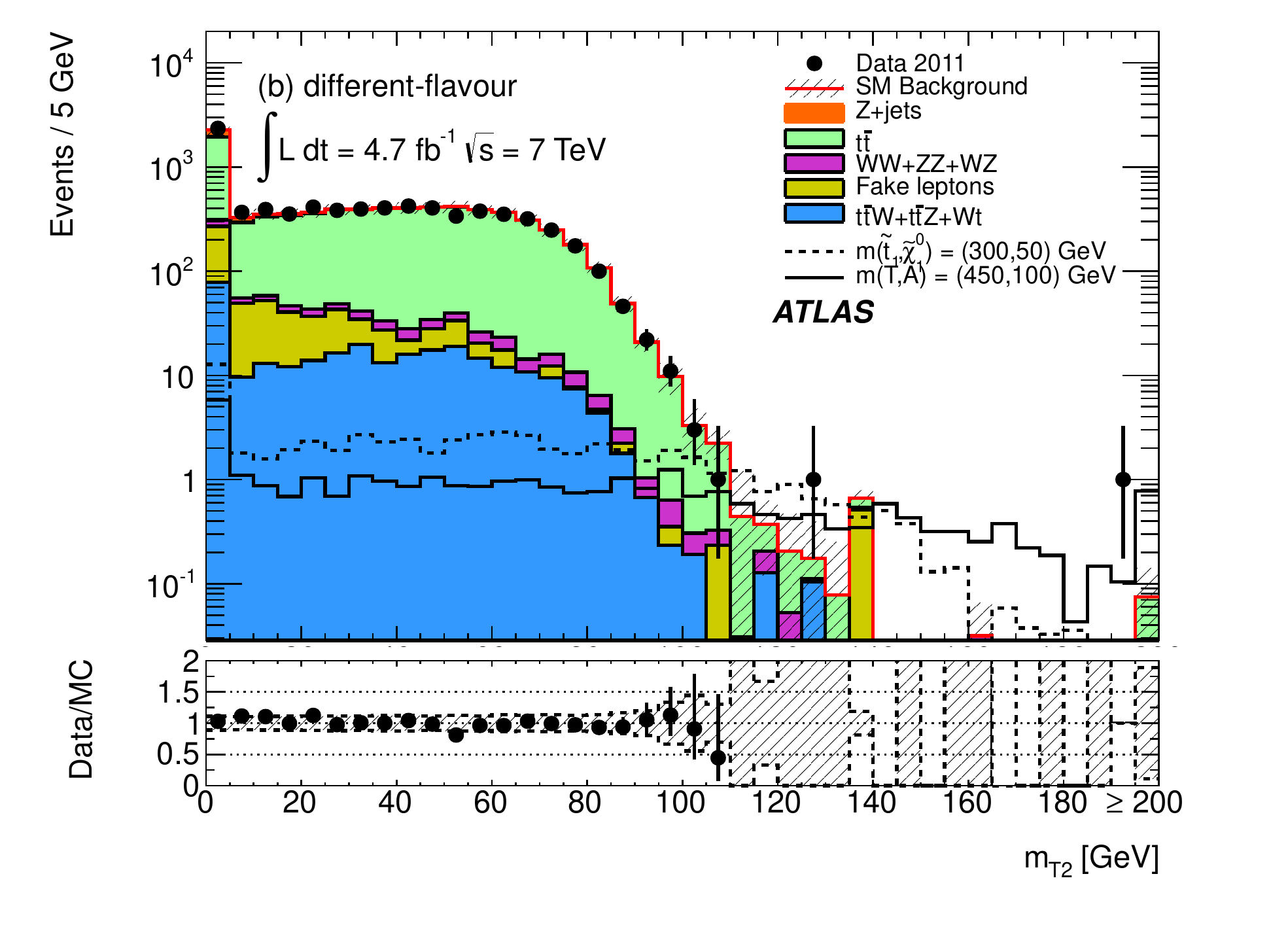}
\end{center}
\caption{Distribution of \mttwo for events passing all the 
signal candidate selection requirements, except that on \mttwo,  
for (a) same-flavour and (b) different-flavour events. 
The contributions from all SM backgrounds are shown; the bands
represent the total uncertainties. The components
labelled ``fake lepton" are estimated from data as described in the text; the 
other backgrounds are estimated from MC simulation with normalizations
measured in control regions described in section~\ref{sec:backgrounds} for \ttbar and 
$Z/\gamma^\star$+jets .
The distributions of the signal expected for two models considered in this paper 
are also shown: the dashed line corresponds 
to signal with a $300 \GeV$ \scalartop and a $50 \GeV$ neutralino,
while the solid line corresponds to a signal with a $450 \GeV$ 
spin-1/2 top quark partner $T$ and a $100 \GeV A_0$ particle.}
\label{fig:mt2}
\end{figure}

Table~\ref{tab:results} shows the expected number of events in the SR for each background 
source and the observed number of events. No excess of events in data is observed, and 
limits at 95\% CL are derived on the visible cross section 
$\sigma_{\mathrm{vis}} = \sigma \times \epsilon \times \mathcal{A}$,
where $\sigma$ is the total production cross section for the non-SM signal, 
$\mathcal{A}$ is the 
acceptance defined by the fraction of events passing the geometric and kinematic selections
at particle level, and $\epsilon$ is the detector reconstruction, identification and trigger
efficiency. Limits are set using the CLs likelihood ratio prescription
as described in ref.~\cite{Read:2002hq}. 
Systematic uncertainties are included in the likelihood function as nuisance parameters 
with a gaussian probability density function. 
The limits are listed in table~\ref{tab:results}.

\begin{table} [htb!]
\caption{Number of expected SM background events and  
number of observed events for the same-flavour (SF) and
different-flavour (DF) signal regions. The quoted errors  are the total uncertainties on
the expected rates. For $\Zboson/\gamma^\star$+jets and $\ttbar$ the 
ratio between the estimate based on the control region and the 
MC prediction (scale factor) is also reported.
Dashes indicate negligible background 
expectations. The expected yield for two signal models is also shown,
with the associated theoretical and experimental uncertainties.
Observed and expected upper limits at 95\% confidence
level on $\sigma_{\mathrm{vis}} = \sigma \times \epsilon \times \mathcal{A}$
are also shown.} 
\label{tab:results}
\begin{center}
\begin{tabular}{|l|c|c|}
\hline
                       & SF         &  DF   \\
\hline
$\Zboson/\gamma^\star$+jets          &   $1.2 \pm 0.5$   &        -          \\
($\Zboson/\gamma^\star$+jets scale factor)   &          (1.27)                  &        -         \\
\hline
$\ttbar$                                                   &    $0.23 \pm 0.23$      &   $0.4 \pm 0.3$ \\
($\ttbar$  scale factor)                                            &              (1.21)             &                (1.10)      \\
\hline
$\ttbarW + \ttbarZ$                            &    $0.11 \pm 0.07$              & $0.19 \pm 0.12$  \\
$WW$                                                 &    $0.01^{+0.02}_{-0.01}$ &  $0.19 \pm 0.18$ \\
$WZ + ZZ$                                         & $0.05 \pm 0.05$            &  $0.03 \pm 0.03$ \\                            
$W t$                                       & $0.00^{+0.17}_{-0.00}$   &  $0.10^{+0.18}_{-0.10}$  \\
Fake leptons                                      & $0.00^{+0.14}_{-0.00}$ & $0.00^{+0.09}_{-0.00}$  \\
\hline
Total SM                                             & $1.6 \pm 0.6$                   & $0.9 \pm 0.6$ \\ 
\hline
Observed                                          &               1                &                 2  \\
\hline 
$m_{\tone} = 300$~\GeV, $m_{\ninoone}$= 50~\GeV 
                                                                                           & $2.2 \pm 0.3 \mbox{(th.)} \pm 0.2 \mbox{(exp.)}$ 
                                                                                            & $3.7 \pm 0.6 \mbox{(th.)} \pm 0.3 \mbox{(exp.)}$ \\
$m_T$ = 450~\GeV, $m_{A_0}$= 100~\GeV & $3.1 \pm 0.4 \mbox{(th.)} \pm 0.3 \mbox{(exp.)}$ 
                                                                                            & $5.8 \pm 0.7 \mbox{(th.)} \pm 0.5 \mbox{(exp.)}$ \\
\hline
95\% CL limit on $\sigma^{\mathrm{obs}}_{\mathrm{vis}} [\mbox{fb}]$  &  $ 0.86$    &   $   1.08$  \\
95\% CL limit on $\sigma^{\mathrm{exp}}_{\mathrm{vis}} [\mbox{fb}]$  &  $ 0.89$    & 0.79 \\    
\hline
\end{tabular}
\end{center}
\end{table}

The results obtained are used to derive limits on the mass of a pair-produced heavy top-quark 
partner decaying into a top quark and a weakly interacting particle with 100\% branching ratio. The limits are derived in the plane defined by the masses of the two particles for two scenarios: a model with a \scalartop $\stop_1$ and a spin-1/2 neutralino $\ninoone$ and one with a spin-1/2 top-quark partner $T$ and a scalar boson $A_0$. 

In both scenarios, the limits are derived after combining the same-flavour and 
different-flavour channels. Uncertainties on the detector response, cross section, luminosity and MC statistics are taken into account. The limits are shown in figure \ref{fig:limits} for the scalar top and spin-1/2 top-quark partner models. Using a signal cross section one standard deviation below the central value, 
a spin-1/2 top-quark partner T with a mass between $300 \GeV$ and $480 \GeV$ (if the 
$A_0$ mass is lower than $100 \GeV$) is excluded at 95\% CL.
The region of the mass plane which is excluded
is smaller for \scalartop production, because of the lower production
cross section. A \scalartop of mass close to 
$300 \GeV$ and a nearly massless neutralino is excluded at 95\% CL.

\begin{figure}[htb!]
\begin{center}
\includegraphics[width=0.7\textwidth]{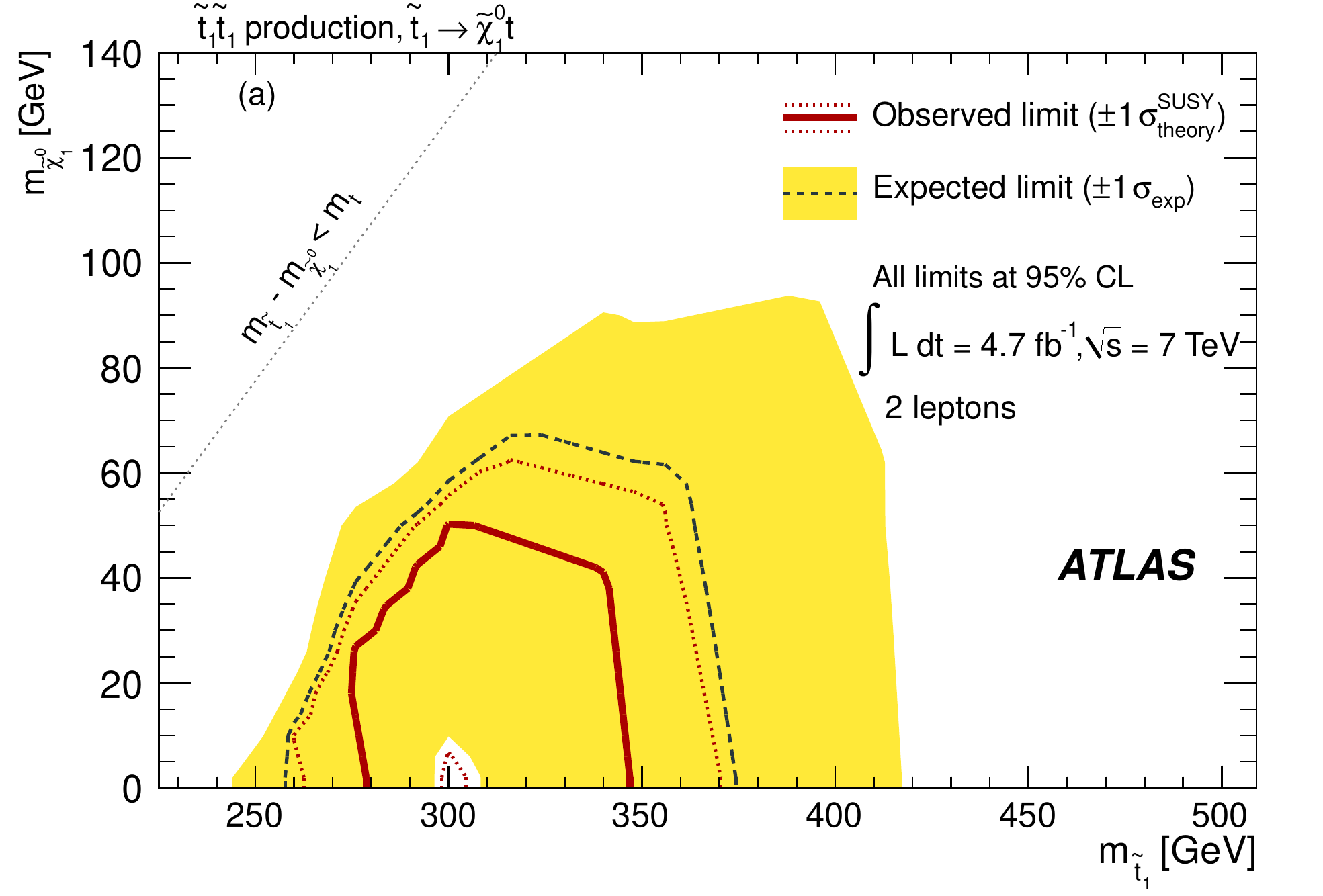}
\includegraphics[width=0.7\textwidth]{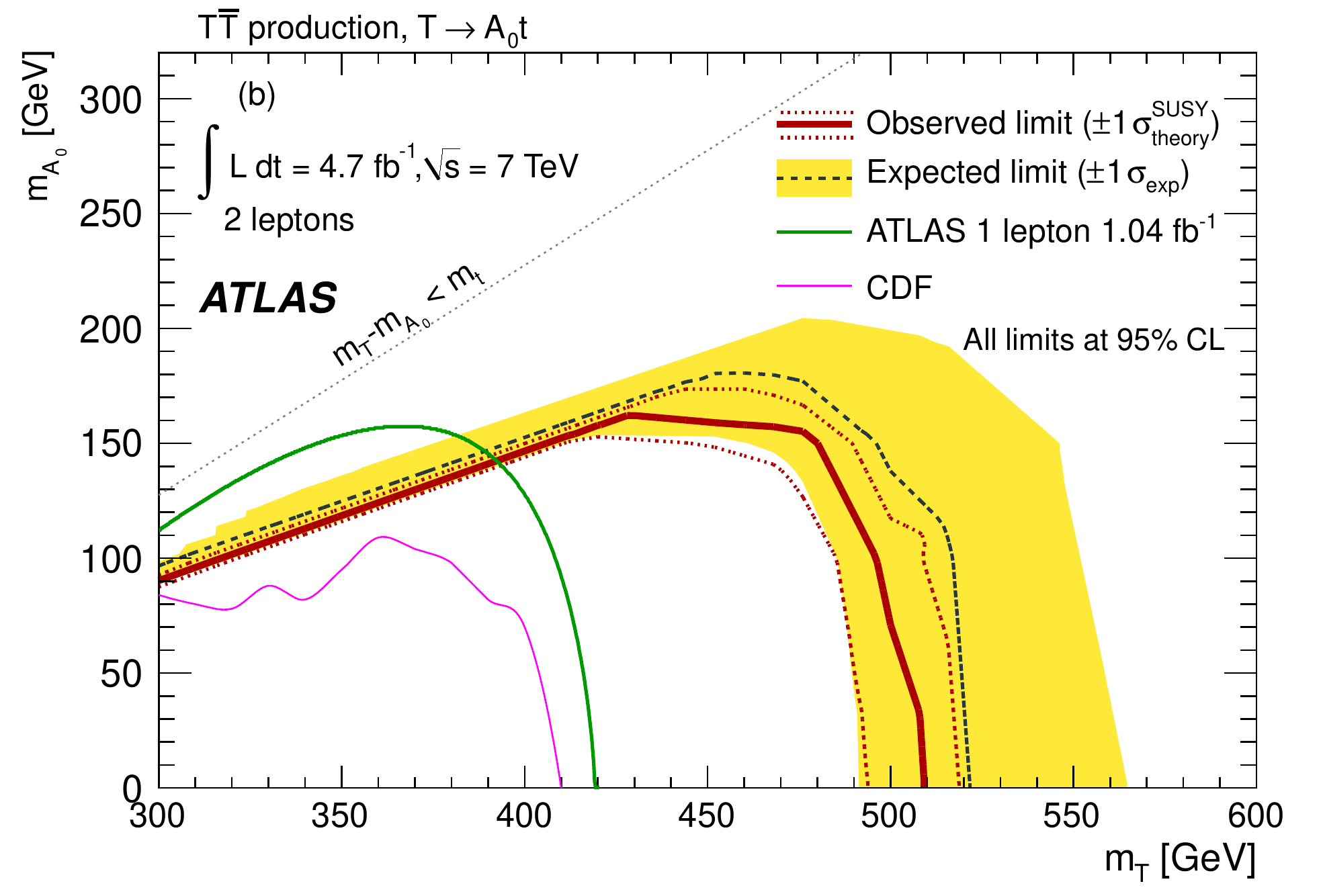}
\end{center}
\caption{Expected and observed 95\% CL limits (a) in the $\tone \rightarrow t \ninoone$ 
model as a function of the \scalartop and neutralino masses, and (b) in the $T \rightarrow t A_0$
model as a function of the spin-1/2 top-quark partner $T$ and $A_0$ masses.
The dashed line and the shaded band are the expected limit and its $\pm1 \sigma$
uncertainty, respectively. The thick solid line is the observed limit for the 
central value of the signal cross section. 
The expected and observed limits do not include the effect of the theoretical 
uncertainties on the signal cross section.
The dotted lines show the effect on the observed limit 
of varying the signal cross section by $\pm 1 \sigma$ 
of the theoretical uncertainty. \label{fig:limits} The curve labelled
``ATLAS 1 lepton 1.04~\ifb" is the previous ATLAS limit from ref.~\cite{Aad:2011wc}
using the one lepton channel 
while the curve labelled ``CDF'' is from ref.~\cite{Aaltonen:2011na}. }
\end{figure}

\section{Conclusions}

A search for a heavy partner of the top quark, which decays into
a top quark and an invisible particle, has been performed using 
4.7~\ifb of pp collision data at $\sqrt{s} = 7  \TeV$ produced by the 
LHC and taken by the ATLAS detector. The 
number of observed events has been 
found to be consistent with the Standard Model expectation.

Limits have been derived on  a spin-1/2 heavy top-quark partner
decaying to a top quark and a heavy neutral particle. A spin-1/2
top-quark partner mass between $300 \GeV$ and $480 \GeV$ is excluded at 95\% CL for 
a heavy neutral particle mass below $100 \GeV$. This result 
extends the previously published limits in this scenario~\cite{Aad:2011wc}.
A supersymmetric \scalartop $\tone$ with a mass of $300 \GeV$
decaying to a top quark and a massless neutralino is also excluded at 95\% CL. 

The present result complements those from other ATLAS direct \scalartop pair production 
searches~\cite{Aad:2011xx,Aad:2011yy,:2012si,:2012ar} addressing different signatures with either both 
\scalartop decaying to a chargino and a {\em b} quark~\cite{Aad:2011xx,Aad:2011yy} 
or with both \scalartop decaying to a neutralino and a top quark and the subsequent top 
quark decays yielding zero or one lepton in the final state~\cite{:2012si,:2012ar}.

\section{Acknowledgements}

We thank CERN for the very successful operation of the LHC, as well as the
support staff from our institutions without whom ATLAS could not be
operated efficiently.

We acknowledge the support of ANPCyT, Argentina; YerPhI, Armenia; ARC,
Australia; BMWF and FWF, Austria; ANAS, Azerbaijan; SSTC, Belarus; CNPq and FAPESP,
Brazil; NSERC, NRC and CFI, Canada; CERN; CONICYT, Chile; CAS, MOST and NSFC,
China; COLCIENCIAS, Colombia; MSMT CR, MPO CR and VSC CR, Czech Republic;
DNRF, DNSRC and Lundbeck Foundation, Denmark; EPLANET and ERC, European Union;
IN2P3-CNRS, CEA-DSM/IRFU, France; GNSF, Georgia; BMBF, DFG, HGF, MPG and AvH
Foundation, Germany; GSRT, Greece; ISF, MINERVA, GIF, DIP and Benoziyo Center,
Israel; INFN, Italy; MEXT and JSPS, Japan; CNRST, Morocco; FOM and NWO,
Netherlands; RCN, Norway; MNiSW, Poland; GRICES and FCT, Portugal; MERYS
(MECTS), Romania; MES of Russia and ROSATOM, Russian Federation; JINR; MSTD,
Serbia; MSSR, Slovakia; ARRS and MVZT, Slovenia; DST/NRF, South Africa;
MICINN, Spain; SRC and Wallenberg Foundation, Sweden; SER, SNSF and Cantons of
Bern and Geneva, Switzerland; NSC, Taiwan; TAEK, Turkey; STFC, the Royal
Society and Leverhulme Trust, United Kingdom; DOE and NSF, United States of
America.

The crucial computing support from all WLCG partners is acknowledged
gratefully, in particular from CERN and the ATLAS Tier-1 facilities at
TRIUMF (Canada), NDGF (Denmark, Norway, Sweden), CC-IN2P3 (France),
KIT/GridKA (Germany), INFN-CNAF (Italy), NL-T1 (Netherlands), PIC (Spain),
ASGC (Taiwan), RAL (UK) and BNL (USA) and in the Tier-2 facilities
worldwide.

\bibliographystyle{atlasnote}

\providecommand{\href}[2]{#2}\begingroup\raggedright\endgroup

\onecolumn

\clearpage 

\begin{flushleft}
{\Large The ATLAS Collaboration}

\bigskip

G.~Aad$^{\rm 48}$,
T.~Abajyan$^{\rm 21}$,
B.~Abbott$^{\rm 111}$,
J.~Abdallah$^{\rm 12}$,
S.~Abdel~Khalek$^{\rm 115}$,
A.A.~Abdelalim$^{\rm 49}$,
O.~Abdinov$^{\rm 11}$,
R.~Aben$^{\rm 105}$,
B.~Abi$^{\rm 112}$,
M.~Abolins$^{\rm 88}$
% ...
\bigskip

% ...
$^{11}$ Institute of Physics, Azerbaijan Academy of Sciences, Baku, Azerbaijan\\
$^{12}$ Institut de F\'{i}sica d'Altes Energies and Departament de F\'{i}sica de la Universitat Aut\`{o}noma de Barcelona and ICREA, Barcelona, Spain\\
% ...
$^{21}$ Physikalisches Institut, University of Bonn, Bonn, Germany\\
% ...
$^{48}$ Fakult\"{a}t f\"{u}r Mathematik und Physik, Albert-Ludwigs-Universit\"{a}t, Freiburg, Germany\\
$^{49}$ Section de Physique, Universit\'{e} de Gen\`{e}ve, Geneva, Switzerland\\
% ...
$^{88}$ Department of Physics and Astronomy, Michigan State University, East Lansing MI, United States of America\\
% ...
$^{105}$ Nikhef National Institute for Subatomic Physics and University of Amsterdam, Amsterdam, Netherlands\\
% ...
$^{111}$ Homer L. Dodge Department of Physics and Astronomy, University of Oklahoma, Norman OK, United States of America\\
$^{112}$ Department of Physics, Oklahoma State University, Stillwater OK, United States of America\\
% ...
$^{115}$ LAL, Universit\'{e} Paris-Sud and CNRS/IN2P3, Orsay, France\\
% ...
\end{flushleft}


\end{document}